\documentstyle[12pt]{article}
\newcommand{\OL}[1]{ \hspace{2pt}\overline{\hspace{-2pt}#1
   \hspace{-1pt}}\hspace{1pt} }
\newcommand{\kten}{{\kappa_{10}}}
\newcommand{\calR}{{\cal R}}
\newcommand{\calG}{{\cal G}}
\newcommand{\tilF}{{\tilde F}}
\newcommand{\barG}{{\OL G}}
\newcommand{\im}{{\rm Im}}

\newcommand{\taub}{{\bar \tau}}
\newcommand{\tilg}{{\tilde g}}
\newcommand{\Mhat}{{\widehat{\cal M}}}
\newcommand{\calB}{{\cal M}}
\newcommand{\calK}{{\cal K}}
\newcommand{\calm}{{\cal M}}
\newcommand{\call}{{\cal L}}
\newcommand{\CN}{{\cal N}}

\renewcommand{\j}{\jmath}

\newcommand{\sect}[1]{\section{#1}\setcounter{equation}{0}}

\newcommand{\leaveout}[1]{}

\begin{document}
\bigskip
\hskip 5in\vbox{\baselineskip12pt
\hbox{NSF-ITP-01-37}
\hbox{SLAC-PUB-8807}
\hbox{SU-ITP-01/16}
\hbox{hep-th/0105097}}
\bigskip\bigskip

\centerline{\Large\bf Hierarchies from Fluxes in String Compactifications}
\bigskip
\bigskip
\bigskip
\centerline{{\bf Steven B. Giddings}$^{1,2}$,  {\bf Shamit Kachru}$^{1,3}$
and {\bf Joseph Polchinski}$^{1}$}
\bigskip
\bigskip
\centerline{$^{1}$Institute for Theoretical Physics, University of
California, Santa Barbara, CA\ 93106-4030}
\smallskip
\centerline{$^{2}$Department of Physics, University of
California, Santa Barbara, CA\ 93106}
\smallskip
\centerline{$^{3}$Department of Physics and SLAC,
Stanford University, Stanford,
CA\ 94305/94309}
\bigskip

\begin{abstract}

Warped compactifications with significant warping provide one of the
few known mechanisms for naturally generating large hierarchies of
physical scales.  We demonstrate that this mechanism is realizable
in string theory, and give examples involving orientifold compactifications
of IIB string theory and F-theory compactifications on Calabi-Yau
four-folds.  In each case, the hierarchy of scales is fixed by a choice of
RR and NS fluxes in the compact manifold. Our solutions involve 
compactifications of the Klebanov-Strassler gravity dual to a confining 
${\cal N}=1$ supersymmetric gauge theory,
and the hierarchy reflects the small scale of chiral symmetry breaking
in the dual gauge theory.

\end{abstract}
\newpage

\baselineskip=17pt

\sect{Introduction}

The origin of the small ratio $M_{\rm weak}/M_{\rm Planck}$ is a great
puzzle.  There are several known mechanisms for producing an exponentially
small ratio of scales.  One is dimensional transmutation, which Nature
employs in many contexts.  Another is nonperturbative effects, such as
instantons, which are exponentially small in the inverse coupling.  A third
possibility has recently come to the fore.  In a warped spacetime --- one
where the normalization of the four-dimensional metric varies in the
transverse dimensions --- a given invariant energy scale can give rise to
many four-dimensional scales, depending on the position-dependent
gravitational redshift in the transverse space. This mechanism has in
particular played a role in the Randall-Sundrum (RS) models~\cite{RS1,RS2}.

Such generation of a hierarchy via redshift
has a number of interesting potential consequences.  For
example, one may reach thresholds to produce Kaluza-Klein modes at low
energies, perhaps in the TeV range, with interesting phenomenological
consequences.  Moreover, in such scenarios, scattering at apparently low
energies can actually reach the fundamental Planck scale, due to the relative
redshift, raising the prospect of experimental probes of Planck- or
string-scale physics at energies far below the apparent
four-dimensional Planck
scale; an example is the possibility of producing black holes at relatively
low energy scales~\cite{GiKa}.

Warped metrics are quite natural in string theory, where D-branes
generically provide sources for the warping.  Within the context of string
compactifications, a particularly simple realization was described by
H. Verlinde~\cite{Verlinde}: simply take $N$ D3-branes to be coincident on
a Calabi-Yau (CY) space.  As is familiar from the AdS/CFT
duality~\cite{Maldacena}, the spacetime near the D3-branes is of the form
$AdS_5 \times S^5$.  It is well known that $AdS_5$ can be represented as a
Poincar\'e-invariant four-dimensional space plus a radial direction, with a
varying warp factor that vanishes at the horizon of its Poincar\'e
parameterization.

The RS models, and the warped compactifications of Verlinde, allow a large
hierarchy but do not explain it.  There is a moduli space of solutions, and the
size of the hierarchy is a function of the moduli.  These moduli correspond, for
example, to the separations of various branes.  Goldberger and Wise~\cite{GW}
have shown that additional dynamics can fix the moduli and produce a calculable
large hierarchy.  Their analysis was phenomenological; the goal of our paper is
to examine this issue in string theory, in the framework suggested by
Verlinde. In particular,
as has been exhibited in the
work of \cite{KS} (see also \cite{GVW,vatop}), a
natural
mechanism to generate such a hierarchy is to consider warped
compactifications with both RR and NS fluxes present.

One way to understand this arises from a picture where branes are placed at
a singularity.
The low energy physics of D3-branes on a CY manifold is conformally
invariant and $\CN=4$ supersymmetric.  In order to fix the moduli it is
necessary to break the conformal invariance and most of the supersymmetry.
Precisely this same issue arises in the context of Maldacena duality.
String theory on $AdS_5 \times S^5$ is dual to $\CN=4$ supersymmetric
Yang-Mills theory.  To find string duals of
gauge theories with confinement and chiral symmetry breaking
one must reduce the symmetry; in the supergravity context this generates
potentials which can fix some of the moduli and stabilize a hierarchy.

A simple means of reducing the symmetry is to place the D3-branes not at a
smooth point of the transverse space but at a
singularity~\cite{kacsil,kehag,klebwit,morpless}.  Indeed, placing them at
a generic CY singularity, a conifold point~\cite{conifold}, reduces
the supersymmetry to $\CN=1$.  This does not break the conformal invariance,
so it is also
necessary to add additional `fractional' branes localized at the
conifold singularity~\cite{gubkleb,klebnek,klebtsey}.
In the final
analysis these branes dissolve into flux, and result in a
nonsingular solution that has recently been found by Klebanov and
Strassler (KS)~\cite{KS}.  So while
the picture of branes and fractional branes at a conifold
is used to motivate the construction, the net result is that one ends up
with a string background with RR and NS fluxes, which lead to a smooth string
solution with a large hierarchy.

The KS solution is, however, noncompact and therefore not suitable as a
means of reducing string theory to four dimensions; in particular it would
produce an infinite 4d Planck scale.  Thus, our goal is to find true string
compactifications, with a finite 4d  Planck scale and a local region
of the KS form which generates a large but finite hierarchy.  This hierarchy
will be determined by the quantized values of the fluxes on the compact
manifold.  (For another discussion of compactifications with fluxes, see 
\cite{Mayr:2000hh}.)

The outline of our paper is as follows.  In section~2 we consider global
constraints on warped IIB solutions.  Such constraints have been used in the
past to exclude warped solutions of IIB supergravity, but in the context of
string theory their effect is to require the presence of objects of negative
tension such as O3 planes and wrapped D7-branes.  Further, when the localized
sources satisfy a certain BPS-like bound, we are able from the global
constraints to find the general solution.  We find that, in the classical
approximation in which we work, the radial modulus is a flat direction with
zero cosmological constant.
This is the case even though supersymmetry is generically broken at
a scale that depends on the radial modulus.  Thus, these are {no-scale
models}~\cite{noscale}.

In section~3 we focus on the local structure of the compactifications,
beginning with a review of the Verlinde solution and its
generalizations.  In particular, in the presence of certain fluxes on a
compact manifold, together with the required O3 planes or D7-branes, we
show that compact smooth string solutions exist with the
hierarchy fixed by the fluxes, in a limit of large fluxes.  However,
as noted above, the overall radius of the compact dimensions is
always left unfixed.  This reflects the familiar feature of string
compactifications, that it is very difficult to stabilize all moduli, though
we should note that in classical IIB compactifications with fluxes the
dilaton generically {\it is} stabilized.\footnote{More
general compactifications with fluxes will be discussed in
ref.~\cite{toappear}.  In particular, some of these have no moduli, and are
reliably studied in a regime where low-energy supergravity is valid.} In
fact the effective theories that we find are very similar to those which
arise in heterotic
string compactifications~\cite{witeff,gluino}.  We also outline the dual,
gauge theory, description of these solutions.  Section~4 is devoted to
constructing explicit examples, first as orientifolds of CY
compactifications, and then as F-theory compactifications (which allow larger
fluxes and hierarchies).

\sect{Warped compactifications: global constraints}

We begin by working in the approximation
of low energy IIB supergravity, with
such localized sources as arise in string theory.  In pure
supergravity, the integrated field equations rule out warped
compactifications under broad conditions~\cite{dewit,maldanun}.  In
section~2.1 we revisit this argument with localized sources included, and
show that a warped compactification is possible if sources with negative
tension are present; such objects do exist in string theory.

With the constraint thus weakened, it does not appear possible to give a
simple description of the general warped solution.  In section~2.2 we show
that when the localized sources satisfy a certain BPS-like bound
involving their energy-momentum tensor and their D3-brane charge, then the
global constraints do determine the general solution.  The localized
sources that we consider --- D3-branes,
wrapped D7-branes, and O3-planes --- all satisfy this bound.  We discuss
various special properties of these solutions, in particular the effective
action for their moduli, and we relate them to solutions recently considered
in the literature.

\subsection{Action, equations of motion, and constraints}

Our starting point is the effective action\footnote{We use the conventions
of \cite{joesmagnumopus}.}
\begin{eqnarray}
S_{\rm IIB} &=& {1\over 2\kten^2} \int d^{10} x \sqrt{-g_{\rm s}}\Biggl\{
e^{-2\phi}
\left[\calR_{\rm s} + 4(\nabla \phi)^2 \right] - {F_{(1)}^2\over 2} -
{1\over 2 \cdot 3!} G_{(3)} \cdot\barG_{(3)} -
{\tilF_{(5)}^2\over 4\cdot5!} \Biggr\} \nonumber\\
&& + {1\over 8 i \kten^2}
\int e^{\phi}{C_{(4)}\wedge
G_{(3)}\wedge\barG_{(3)}}\ +\ S_{\rm loc} \ .
\label{IIBS}
\end{eqnarray}
Here $g_{\rm s}$ denotes the string metric.  We have also defined the
combined three-flux, $G_{(3)}= F_{(3)} - \tau H_{(3)}$, where as usual $\tau=
C_{(0)} + i e^{-\phi}$, and
\begin{equation}
{\tilde F}_{(5)} = F_{(5)} - {1\over 2} C_{(2)}\wedge H_{(3)} + {1\over 2}
B_{(2)}\wedge F_{(3)}\ .
\end{equation}
The term $S_{\rm loc}$ is the action of localized
objects, such as branes, which will
become important shortly.  The condition
${\tilde F}_{(5)} = {*}{\tilde F}_{(5)}$ must as usual be
imposed by hand on the equations of motion.

We will be considering compactifications arising from F-theory, so it is
particularly useful to reformulate the action in an $SL(2,{\bf Z})$ invariant
form by defining the Einstein metric $g_{MN} = e^{-\phi/2} g_{{\rm
s}\,MN}$, whence the action becomes
\begin{eqnarray}
S_{\rm IIB} &=& {1\over 2\kten^2} \int d^{10}x \sqrt{-g} \Biggl\{ \calR -
{\partial_M\tau \partial^M \taub \over 2 ({\rm Im}\, \tau)^2} -
 {G_{(3)}\cdot\barG_{(3)} \over
12\, {\rm Im}\, \tau }- {\tilF_{(5)}^2\over 4\cdot 5!}\Biggr\}
\nonumber\\
&&\qquad\qquad\qquad\qquad + {1\over
8i\kten^2}
\int {C_{(4)}\wedge G_{(3)}\wedge \barG_{(3)} \over {\rm Im}\, \tau}\ +\
S_{\rm loc}\ .
\label{IIBE}
\end{eqnarray}
Henceforth we use the Einstein metric throughout.  Invariance under the
$SL(2,{\bf Z})$ transform
\begin{equation}
\tau\rightarrow {a\tau+b \over c\tau+d }\ ,
\label{monod}
\end{equation}
together with the transformation
\begin{equation}
G_{(3)}\rightarrow {G_{(3)} \over  c\tau +d}
\label{monodg}
\end{equation}
is readily checked.

Our interest is in warped metrics maintaining four-dimensional Poincar\'e
symmetry, with convenient parameterization
\begin{equation}
ds_{10}^2 = e^{2A(y)} \eta_{\mu\nu} dx^\mu dx^\nu + e^{-2A(y)} \tilg_{mn}
dy^m dy^n
\label{mansatz}
\end{equation}
in terms of four-dimensional coordinates $x^\mu$ and coordinates
$y^m$ on the compact manifold $\calm_6$.
The axion/dilaton will be allowed to vary over the compact manifold,
\begin{equation}
\tau=\tau(y)\ ,
\end{equation}
and since we will consider D7-branes, monodromies of the form
(\ref{monod}) will be allowed.
To maintain Poincar\'e invariance only compact components of $G_{(3)}$ are
present, and furthermore, with monodromies (\ref{monodg}),
these will transform in a
non-trivial bundle over $\calm_6$:
\begin{equation}
G_{(3)}\in \sigma ( \Omega^3 \otimes \call)\ ,
\label{Gbundle}
\end{equation}
where $\Omega$ denotes the canonical bundle, and
$\call$ is the line bundle defined by the transformation law
(\ref{monodg}).
Finally, Poincar\'e invariance and the Bianchi identity
allows a five-form flux of the form
\begin{equation}
\tilF_{(5)} =  (1 +
{*})[d\alpha \wedge dx^0 \wedge dx^1 \wedge dx^2 \wedge dx^3]
\ ,\label{f5ans}
\end{equation}
with $\alpha$ a function on the compact space.
Also, in accord with Poincar\'e invariance, we will allow some number of
D3-branes along the noncompact directions, as well as D7-branes filling
the noncompact directions and wrapping certain four-cycles in $\calm_6$.

Einstein's equation, trace reversed, is
\begin{equation}
\calR_{MN} = \kten^2 \biggl( T_{MN} -\frac{1}{8} g_{MN} T \biggr)\ ,
\end{equation}
where $T_{MN} = T^{\rm sugra}_{MN} + T^{\rm loc}_{MN}$ is the total stress
tensor of the supergravity fields and the localized objects.  In particular,
the latter contribution is
\begin{equation}
T_{MN}^{\rm loc}= -{2\over \sqrt{-g} } {\delta S_{\rm loc} \over \delta
g^{MN} }\ .
\label{tmn}
\end{equation}
The noncompact components take the form
\begin{equation}
\calR_{\mu\nu} = - g_{\mu\nu} \Biggl(
{G_{mnp}\barG^{mnp} \over 48 \,\im\,\tau}
+ {e^{-8A} \over 4} \partial_m\alpha\partial^m\alpha \Biggr) +
\kten^2
\biggl( T_{\mu\nu}^{\rm loc} - {1\over 8} g_{\mu\nu} T^{\rm loc}\biggr)\ .
\label{fourdeq}
\end{equation}
{}From the metric Ansatz (\ref{mansatz}), one computes the Ricci components
\begin{equation}
\calR_{\mu\nu} = - \eta_{\mu\nu} e^{4A} \tilde \nabla^2 A
= - \frac{1}{4} \eta_{\mu\nu} \Bigl( \tilde \nabla^2 e^{4A} -
e^{-4A} \partial_m e^{4A} \partial^{\tilde m} e^{4A}
\Bigr)
\ .
\end{equation}
(A tilde denotes use of the metric $\tilde g_{mn}$.)
Using this and tracing (\ref{fourdeq}) gives
\begin{equation}
\tilde \nabla^2 A  = e^{-2A}
{G_{mnp}\barG^{mnp} \over 48 \,\im\,\tau}
+ \frac{e^{-6A}}{4} \partial_m\alpha\partial^m\alpha
 + {\kten^2 \over 8} e^{-2A} (T_m^m - T_\mu^\mu )^{\rm loc}
\ .
\label{warpeq}
\end{equation}
or
\begin{equation}
\tilde \nabla^2 e^{4A}  = e^{2A}
{G_{mnp}\barG^{mnp} \over 12 \,\im\,\tau}
+ {e^{-6A}} \Bigl[\partial_m\alpha\partial^m\alpha
+ \partial_m e^{4A} \partial^{ m} e^{4A}\Bigr]
 + {\kten^2 \over 2} e^{2A} (T_m^m - T_\mu^\mu )^{\rm loc}
\ .
\label{intconst}
\end{equation}

These equations serve as stringent constraints on flux/brane
configurations that can lead to warped solutions on {\it compact}
manifolds.\footnote{One reaches the same conclusions by considering $
\nabla^2 e^{kA}$ for any positive $k$, but $k=4$ is the value that will be
useful in the next subsection.}
To see this, note that the integrals
of their left sides over a compact
manifold $\calm_6$ vanish, whereas the flux and warp terms on the
right-hand side are positive definite.  Thus, in the absence of localized
sources there is a no-go theorem~\cite{dewit,maldanun}:
the fluxes must vanish and the warp
factor must be constant.  For a warped solution
the stress terms on the RHS must be negative,
which can only be true under certain
circumstances.

For example, consider a $p$-brane wrapped on a $(p-3)$-cycle $\Sigma$
of the manifold
$\calm_6$.  To leading order in
$\alpha'$ (and in the case of vanishing fluxes along the brane) this
contributes a source action
\begin{equation}
S_{\rm loc}= - \int_{R^{4}\times \Sigma} d^{p+1}\xi\, T_p
\sqrt{-{}g} + \mu_p \int_{R^{4}\times \Sigma} C_{p+1}\ ;
\label{braneact}
\end{equation}
for positive tension objects the Einstein frame tension is
\begin{equation}
T_p= |\mu_p| e^{(p-3)\phi/4}\ .
\end{equation}
Eq.~(\ref{braneact}) gives a stress tensor
\begin{equation}
T^{\rm loc}_{\mu\nu} = -T_p e^{2A} \eta_{\mu\nu}\delta(\Sigma) \ ,\quad
T^{\rm loc}_{mn} = -T_p  \Pi_{mn}^\Sigma \delta(\Sigma)\ ,
\label{pstress}
\end{equation}
where $\delta(\Sigma)$ and $\Pi^\Sigma$ denote the delta function and
projector on the cycle $\Sigma$, respectively.
{}From this we find
\begin{equation}
(T_m^m - T_\mu^\mu )^{\rm loc} = (7-p) T_p \delta(\Sigma)\ .
\label{ptrace}
\end{equation}
Eq.~(\ref{ptrace}) tells us that for $p<7$,
in order to have the required negative
stress on the RHS of the constraint (\ref{intconst}), the compactification
must involve {\it negative} tension objects.

String theory does have such objects, and so evades the no-go theorem
of \cite{dewit,maldanun}.  O3 planes are a simple example.  The $T^6/{\bf
Z}_2$ orientifold, which is $T$-dual to the type~I theory,
is a compact Minkowski solution with 16 D3-branes and 64
O3-planes~\cite{joesmagnumopus}.  This implies that the O3 tension
is $-\frac{1}{4}T_3$.  This orientifold was discussed in
ref.~\cite{Verlinde} as an example of a warped string solution.

Note that F-theory compactifications, despite having D7-branes,
dilaton gradients, and RR 1-form fluxes, satisfy the
constraint~(\ref{intconst}) {\it without} negative tension.  This is because
terms involving $\tau$ gradients do not enter the constraint, and
the  D7 brane  stress tensor contributions vanish by
eq.~(\ref{ptrace}).

To be precise, this is true only to leading order in
$\alpha'$.  It is necessary to include also the first $\alpha'$ corrections
to the D7 action $S_{\rm loc}$ (we will explain this expansion below).
In the Chern-Simons action the correction is~\cite{BSV}
\begin{equation}
-\mu_3 \int_{R^4\times\Sigma} C_{(4)} \wedge \frac{p_1({\cal R})}{48}
= \frac{\mu_7}{96} (2\pi\alpha')^2  \int_{R^4\times\Sigma} C_{(4)} \wedge
{\rm Tr}\,(\calR_{(2)} \wedge \calR_{(2)}) \ . \label{CScurr}
\label{cseq}
\end{equation}
This Chern-Simons coupling
captures the induced D3 charge on the wrapped D7-brane.
In the DBI action it is~\cite{dbicorr}
\begin{equation}
-\frac{\mu_7}{96} (2\pi\alpha')^2  \int_{R^4\times\Sigma}
d^{4}x
\sqrt{-{}g} {\rm Tr}
(\calR_{(2)} \wedge {*}\calR_{(2)}) \ . \label{DBIcurr}
\label{dbieq}
\end{equation}
This term computes the first $\alpha^\prime$ correction to the wrapped
D7-brane tension.\footnote{For
simplicity we are considering in eqs.~(\ref{cseq},\ \ref{dbieq}) the case of
a trivial normal bundle; the full form is given in ref.~\cite{norm}.  The
F-theory result~(\ref{Qback}) is general.}
The Chern-Simons coupling has the effect, for example, that a D7-brane
wrapped on K3 has $-1$ unit of D3 charge~\cite{BSV}.  This state is still
BPS, with the same supersymmetry as the D3-brane, so the DBI coupling
must contribute $-T_3$ to the tension.
In F-theory, this background charge
is given in terms of the Euler number of the
corresponding fourfold by
\begin{equation}
Q_3^{\rm D7}= - {\chi(X) \over 24}\ ,
\label{Qback}
\end{equation}
and ${\cal N}=1$ supersymmetry implies the corresponding tension $Q_3^{\rm
loc}
T_3$.  This can be thought of as coming from the summed contribution
of all 7-branes wrapping four-cycles in the base of the elliptic
fibration $X$.  To directly derive this tension along the
lines discussed above, one should
use the generalization of (\ref{dbieq}) which is applicable to
branes wrapping divisors in the (non-CY) base of $X$;
the result is guaranteed by the supersymmetry of the configuration,
and the direct calculation is beyond
the scope of our work.

We have been discussing constraints from the integrated Einstein equation.
The Bianchi identity/equations of motion for the 5-form flux is\footnote{
Recall that $2 \kten^2 = (2\pi)^7 \alpha'^4$,
$\mu_3 = (2\pi)^{-3} \alpha'^{-2}$, $\mu_7 = (2\pi)^{-7}
\alpha'^{-4}$, and, in Einstein frame, $T_3 = \mu_3$~\cite{joesmagnumopus}.}
\begin{equation}
d {\tilde F}_{(5)} = H_{(3)}\wedge F_{(3)} + 2 \kten^2 T_3 \rho_3^{\rm loc}
\label{bi5}
\end{equation}
where $\rho_3^{\rm loc}$ is the D3 charge density form from localized
sources; this includes the contributions of the D7-branes or O3 planes, and
also of mobile D3-branes that may be present.\footnote{In deriving this
field equation there is an annoying subtlety due to the self-dual flux.
The electric coupling of $C_{(4)}$ must actually be half of what we have
written in eqs.~(\ref{braneact},\ \ref{cseq}), in order to obtain
eq.~(\ref{bi5}).  However, any object carrying D3 charge also has a
magnetic coupling to $C_{(4)}$; in a self-dual background the action for a
probe is then obtained by doubling the electric coupling as we have done.

An alternative approach to the self-dual flux is to use a
Lorentz-noninvariant action: double the $F_{(5)}^2$ and Chern-Simons terms
in the actions~(\ref{IIBS},\ \ref{IIBE}) but restrict to terms in which
$F_{(5)}$ or $C_{(4)}$ has a 1-component.  This action, derived by
$T$-duality from the IIA action, is well-suited to the study of
compactification of the IIB theory.}
The integrated Bianchi identity
\begin{equation}
\frac{1}{ 2 \kten^2 T_3}
\int_{\calm_6}  H_{(3)}\wedge F_{(3)}\ +\  Q_3^{\rm
loc}=0
\label{threeconst}
\end{equation}
states that the total D3 charge from supergravity backgrounds and localized
sources vanishes.
In the next subsection, we will analyze the
constraints~(\ref{intconst},\,\ref{threeconst}) further.

Finally, let us discuss  the nature of the $\alpha'$ expansion.
The localized source in the Bianchi identity~(\ref{bi5}) is of order $N
\alpha'^2$, where $N$ is the characteristic D3 charge.  It is not possible
to take $N$ to be parametrically large, because the negative contributions
to the Bianchi identity are determined by the topology of the manifold.
However, the Euler number~(\ref{Qback}) can be quite large in a given
example, and so we will treat $N$ as an effective large parameter as in
ref.~\cite{Verlinde}.  We will then treat $N \alpha'^2$ as being of order
one, but drop order
$\alpha'$ effects such as the string corrections to the supergravity action.
This is why we needed to keep the curvature terms in the D7-brane action.
The Bianchi identity then implies that $G_{(3)} = O(N^{1/2} \alpha')$;
the factor of $\alpha'$ is consistent with the quantization
\begin{equation}
\frac{1}{2\pi\alpha'} \int F_{(3)} \in 2\pi{\bf Z}\ ,\quad
\frac{1}{2\pi\alpha'} \int H_{(3)} \in 2\pi{\bf Z}\ ,\label{3quant}
\label{fluxquant}
\end{equation}
and the number of 3-form flux units then scales as $N^{1/2}$.

\subsection{Special solutions}

\subsubsection{A BPS-like condition}
With general negative tension sources, the constraints from the integrated
field equations appear to be rather weak.  However, in the special case that
\begin{equation}
\frac{1}{4}(T_m^m - T_\mu^\mu)^{\rm loc} \geq T_3\rho_3^{\rm loc}
\label{bps?}
\end{equation}
for all localized sources, the global constraints determine the form of the
solution completely.

In fact, the inequality~(\ref{bps?}) holds for all of
the localized sources considered in this paper.  For D3-branes and O3 planes,
whose integrated $\rho_3$ is respectively $+1$ and $-\frac{1}{4}$, the
stress tensor is
\begin{equation}
T_0^0 = T_1^1 = T_2^2 = T_3^3 = -T_3\rho_3\ ,\quad T_m^m = 0\ ,
\end{equation}
and so the inequality is actually saturated.  Anti-D3-branes satisfy the
inequality but do not saturate it.  D5-branes wrapped on collapsed
cycles also satisfy the inequality, as their tension comes entirely from
their induced D3 charge.

For D7-branes, the
nonvanishing contributions to the two sides of the inequality come from the
curvature terms~(\ref{CScurr},\,\ref{DBIcurr}).  In the simple case of
D7-branes wrapped on K3, the property ${*}\calR_{(2)} = \calR_{(2)}$
implies that the inequality is saturated.  If a
nontrivial gauge bundle is introduced, the inequality is still respected
as a consequence of $F_{\mu\nu} F^{\mu\nu} \geq
F_{\mu\nu} ({*}F)^{\mu\nu}$.  For the more general wrappings that arise in
F-theory, we argue below that the inequality is saturated.

There are objects that do violate the
inequality~(\ref{bps?}).  O5 planes make
a negative contribution to the LHS and zero contribution to the RHS.  Anti-O3
planes make a negative contribution to the LHS and a positive contribution
to the RHS.

The inequality~(\ref{bps?}) resembles a BPS condition.  Indeed, the
underlying IIB supersymmetry algebra implies that
\begin{equation}
H \geq T_3 Q_3\ .\label{bps}
\end{equation}
If this holds locally, as might be expected classically, then by applying
Lorentz invariance we get
$- T_0^0 = -T_1^1 = -T_2^2 = -T_3^3 \geq T_3\rho_3$.  When the
inequality~(\ref{bps}) is saturated, the pressure $T_m^m$ should vanish by
analogy to the no-force condition.  Away from extremality $T_m^m - T_\mu^\mu$
generally increases, by analogy to the weak energy
condition, so the inequality~(\ref{bps?}) follows.
The O planes that do not satisfy the bound~(\ref{bps?}) are able to evade it
because the necessary supercharges do not exist: they are projected out by
the orientifold.  The D7-branes that arise in F-theory compactifications
saturate the bound because they preserve an ${\cal N}=1$ supersymmetry
that is also preserved by D3-branes.

\subsubsection{Solution of the constraints}

In terms
of the potential $\alpha$ the Bianchi identity~(\ref{bi5}) becomes
\begin{equation}
\tilde\nabla^2 \alpha = i e^{2A}
{G_{mnp}({*}_6\barG^{mnp}) \over 12 \,\im\,\tau}
+ 2{e^{-6A}}\partial_m\alpha \partial^{ m} e^{4A}
 + 2\kten^2 e^{2A} T_3 \rho_3^{\rm loc}
\ , \label{lap}
\end{equation}
where ${*}_6$ is the dual in the transverse directions.
Subtracting this from the Einstein equation constraint (\ref{intconst})
gives
\begin{equation}
\tilde \nabla^2 (e^{4A} -\alpha)  =
{e^{2A} \over 6 \,\im\,\tau}  \Bigl|i G_{(3)} - {*}_6G_{(3)}\Bigr|^2
+ {e^{-6A}} |\partial(e^{4A} - \alpha)|^2
 + 2{\kten^2} e^{2A}
\biggl[ \frac{1}{4} (T_m^m - T_\mu^\mu )^{\rm loc} - T_3 \rho_3^{\rm loc}
\biggr]
\ .
\label{subconst}
\end{equation}
The LHS integrates to zero, while under the assumption~(\ref{bps?}) the RHS
is nonnegative.  Thus, if the inequality~(\ref{bps?}) holds, then
\begin{itemize}
\item
The 3-form field strength is imaginary self-dual,
\begin{equation}
{*}_6G_{(3)} = i G_{(3)}\ . \label{imsd}
\end{equation}
\item
The warp-factor and 4-form potential are related,
\begin{equation}
e^{4A} = \alpha\ . \label{warp4}
\end{equation}
\item
The inequality~(\ref{bps?}) is actually saturated.
\end{itemize}

Assuming this form, let use review the full set of field equations and
Bianchi identities.  The 5-form field strength~(\ref{f5ans}) is self-dual by
construction. Its field strength/Bianchi identity~(\ref{lap}) is consistent
and determines $\alpha$ and $A$, provided that the total D3
charge~(\ref{threeconst}) vanishes.  The 3-form Bianchi identities
\begin{equation}
dF_{(3)} = dH_{(3)} = 0 \label{bi3}
\end{equation}
must be imposed.  Using these, the
equation of motion then takes the form
\begin{equation}
d \Lambda + \frac{i}{\im \,
\tau} d\tau\wedge {\rm Re}\,\Lambda = 0\ , \quad \Lambda =
e^{4A} {*}_6G_{(3)} - i \alpha G_{(3)}\ ,
\end{equation}
and so is satisfied as a consequence of eqs.~(\ref{imsd},\,\ref{warp4}).
The
$\calR_{\mu\nu}$ equation also follows from these conditions.  Finally, the
remaining field equations reduce to
\begin{eqnarray}
\tilde \calR_{mn} &=& {\kten^2}
\frac{\partial_{m}\tau \partial_{n} \bar\tau +
\partial_{n}\tau \partial_{m} \bar\tau}{4(\im \, \tau)^2}
+ \kten^2 \biggl( \tilde T^{\rm D7}_{mn} -\frac{1}{8} \tilde g_{mn} \tilde
T^{\rm D7}
\biggr)
\  , \label{tilrmn}\\
\tilde\nabla^2\tau &=& {\tilde\nabla \tau\cdot \tilde\nabla\tau \over
i \,\im\,\tau} - \frac{4\kappa_{10}^2(\im \, \tau)^2}
{\sqrt{-g}}
{\delta \tilde S_{\rm D7}
\over \delta\taub }\ .
\label{taueqn}
\end{eqnarray}
These are the equations determining a solution to F-theory in the
supergravity approximation.

In summary, assuming that the localized sources satisfy~(\ref{bps?}),
the necessary and sufficient conditions for a solution are an underlying
manifold $\widetilde{\cal M}_6 \equiv (\tilde g_{mn},\tau)$
satisfying~(\ref{tilrmn},\,\ref{taueqn}), closed 3-form fluxes
$F_{(3)}$ and $H_{(3)}$ such that $G_{(3)}$ is imaginary self-dual, and
vanishing total D3 charge.

\subsubsection{Supersymmetry, and relation to previous solutions}

The conditions for ${\cal N}=1$ supersymmetry of such a solution have recently
been considered in refs.~\cite{GP,gubs} for constant dilaton, and in
ref.~\cite{GP2} more generally.  The underlying manifold must be K\"ahler
and the connection
$\tilde D_{m} -{i\over2} Q_m$ must lie in $SU(3)$, where $Q_m$ is constructed
from $\tau$ as in~\cite{Schwarz}.  The flux $G_{(3)}$ must be a (2,1) form
and primitive, meaning that the index structure is $\bar\imath jk$ and
the contraction with the K\"ahler form $J^{\bar\imath j}$ vanishes.
The condition ${*}_6G_{(3)} = i G_{(3)}$ allows a primitive $(2,1)$ piece
and a $(0,3)$ piece.\footnote{It also allows a
$(1,2)$ piece of the form $K_{(2)} \wedge \omega_{(1)}$ where
$K_{(2)}$ is the
K\"ahler form and $\omega_{(1)}$ is a nontrivial closed $(0,1)$-form.
A compact Calabi-Yau manifold has no such $(0,1)$-form, and neither
do the Calabi-Yau orientifolds or F-theory compactifications we
consider.  Note that in our conventions for the complex
basis, $\epsilon_{123}{}^{123}=-i$.}
Thus our general solution is supersymmetric if and only if the $(0,3)$ part
vanishes.

In general, supersymmetric and nonsupersymmetric solutions are both
possible, though the latter are more generic.  Consider for example the
$T^6/{\bf Z}_2$ orientifold.  This is somewhat
special because it has ${\cal
N}=4$ supersymmetry in the absence of
$G$-flux, but it serves for illustration.  In terms of three complex
coordinates, the primitive fluxes $G_{\bar123}$, $G_{1\bar23}$, and
$G_{12\bar3}$ can be turned on consistent with the quantization
conditions~(\ref{3quant}) (these fix $\tau$ and some of the K\"ahler
moduli), leaving ${\cal N}=1$ supersymmetry.  If the additional flux
$G_{\bar1\bar2\bar3}$ is nonzero then all supersymmetry is broken.

Noncompact solutions of this form have previously been described in
ref.~\cite{GP} in the special case of constant dilaton.  The supersymmetric
solutions are dual~\cite{GVW,DGS} to the M theory solutions of
ref.~\cite{beck2}.  As emphasized in ref.~\cite{GP2} these solutions are
special, in the sense that the ${\cal N}=1$ supersymmetry lies in an ${\cal
N}=4$ subgroup of the full ${\cal N}=8$ IIB supersymmetry.  In IIB form,
this is the subgroup preserved by a space-filling D3-brane; in M theory
form it is the subgroup preserved by a space-filling M2-brane.
F-theory compactifications on  CY fourfolds preserve ${\cal N}=1$
supersymmetry in the presence of D3 branes (and in fact are limits
of the M theory solutions of \cite{beck2}).
Therefore, we can infer that they are
solutions of this special form, though we have
not displayed this by computing and explicitly comparing
the contributions of (the fully generalized forms of)
(\ref{dbieq}) and (\ref{cseq}) for the wrapped 7-branes.

\subsubsection{Moduli and effective actions}

The necessary and sufficient
conditions
(\ref{threeconst},\,\ref{imsd},\,\ref{bi3},\,\ref{tilrmn},\,\ref{taueqn}) are
all invariant under rescaling $\tilde g_{mn} \to \lambda^2
\tilde g_{mn}$.  Thus,
\begin{itemize}
\item
All special solutions have a radial modulus.
\end{itemize}
Thus our goal of fixing the moduli in a warped compactification is limited
in this class of solutions to leaving at least this one.  On the other hand,
there is no dilaton modulus, because the dilaton couples differently to the
NS-NS and R-R 3-form fluxes and so has a nontrivial potential.  This
suggests that it may be an interesting exercise to look for solutions having
no classical moduli by introducing sources not satisfying the
inequality~(\ref{bps?})~\cite{toappear}.

This is slightly subtle, because the solution itself does not scale simply.
In  the field equation~(\ref{intconst}), the terms involving derivatives of
$A$ scale like $\lambda^{-2}$, and the flux source term scales like
$\lambda^{-6}$. 
It follows that at large radius
$e^{4A} = 1 + O(\lambda^{-4})$ and so the warp factor approaches a
constant.  At radii less than $O(N^{1/4} \alpha'^{1/2})$ the warping becomes
significant.

The properties of the nonsupersymmetric solutions --- vanishing
four-dimensional cosmological constant and a radial modulus in spite of the
absence of supersymmetry --- identifies them as {\it no-scale
models}~\cite{noscale,witeff,gluino}.  The combination of broken
supersymmetry with vanishing cosmological constant is intriguing, but there
is no known reason that it should survive quantum corrections, from
instantons and even perturbative loops.  Even at string tree level,
$\alpha'$ corrections to the supergravity field equations presumably spoil
the no-scale structure.

Let us also consider the effective four-dimensional action.
Before turning on fluxes, there will be massless fields corresponding to
the K\"ahler and complex structure moduli; we denote the latter
$z^\alpha$.  Furthermore, for orientifold
models, the dilaton field $\tau$ is massless, whereas in general F-theory
models it is fixed in terms of the complex structure moduli by
(\ref{taueqn}).  For the moment we consider the case of a single K\"ahler
modulus, the radial modulus, in a four-dimensional superfield $\rho$.

For a large-radius CY or orientifold, the K\"ahler potential follows by
dimensionally reducing the 10d action.\footnote{For
further discussion see the appendix.}  For the radius we find
\begin{equation}
\calK(\rho)=-3\ln[-i(\rho-{\bar \rho})]\ ,
\end{equation}
and for the dilaton and complex structure moduli
\begin{equation}
\calK(\tau,z^\alpha) = -\ln[-i(\tau-{\bar \tau})] - \ln\left(-i\int_{\cal M}
\Omega\wedge {\bar \Omega} \right)
\label{KWP}
\end{equation}
where $\Omega$ is the holomorphic $(3,0)$ form.  The latter expression
follows from the Weil-Petersson metric, and is discussed in \cite{modsp}.
An obvious conjecture for the F-theory generalization of (\ref{KWP}) is
\begin{equation}
 \calK = -\ln\left(\int_{ X}
\Omega_4\wedge {\bar \Omega_4} \right)
\end{equation}
where $X$ and $\Omega_4$ denote the CY fourfold and its holomorphic
$(4,0)$ form respectively.

The fluxes generate a superpotential, which takes the form \cite{GVW}
\begin{equation}
W = \int_{\cal M} \Omega \wedge G_{(3)}\ .
\end{equation}
This
is independent of $\rho$.
The expected F-theory generalization of this formula takes the form
\cite{GVW}
\begin{equation}
W=\int_{X} \Omega_4 \wedge G_{(4)}\ .
\label{Fsuper}
\end{equation}
In (\ref{Fsuper}), $G_{(4)}$ denotes the four-form flux one would
get in M-theory by compactifying the F-theory on a circle; it
can be expressed in terms of type IIB quantities in the F-theory
limit.
If the one works with a local trivialization of the elliptic
fibration, for example in the vicinity of the conifold point, with
fiber coordinate $w$,
the four form $G_{(4)}$ takes the form
\begin{equation}
G_{(4)}= -{G_{(3)} d{\bar w}\over \tau-{\bar \tau}} + {\rm h.c.}\ .
\label{Gfdef}
\end{equation}
We will further discuss issues surrounding use of such a trivialization in
section four.

Under these conditions the ${\cal N}=1$ supergravity potential
simplifies~\cite{noscale},
\begin{eqnarray}
{\cal V} &=& \frac{1}{2\kappa_{10}^2} e^{\calK} \Bigl(
G^{a\bar b} D_a W \OL{D_{ b} W}
 - 3 |W|^2 \Bigr)
\nonumber\\
&\to&
\frac{1}{2\kappa_{10}^2} e^{\calK} \Bigl(  G^{ i\bar \j}
D_i W \OL{D_{j}
W}
\Bigr)
\label{pot}\ ,
\end{eqnarray}
where $D_a W = \partial_a W + W \partial_a \calK$ 
and $G_{a\bar b} = \partial_a
\partial_{\bar b} \calK$, and the indices $a,b$ are summed over superfields,
with $i,j$ labeling indices excluding $\rho$.
In no-scale models the $|D_\rho W|^2$ term cancels the negative
term, leaving a nonnegative potential.  When $D_a W = 0$ the potential
vanishes; this condition is independent of $\rho$, so if there are $n$
superfields besides
$\rho$ it represents $n$ equations on $n$ moduli and leaves $\rho$
undetermined.  Generically at these solutions
$W
\neq 0$, so $D_\rho W = -3 W/(\rho-\OL \rho)$ is
nonzero and supersymmetry is
broken.

A useful check on these expressions comes by comparing the 4d and 10d
equations.  In the CY/orientifold case, one readily finds (see appendix)
\begin{eqnarray}
0=D_\alpha W &\equiv& \partial_\alpha W + (\partial_\alpha 
\calK) W = \int_{\cal
M} G_{(3)}
\wedge
\chi_\alpha \ ,\nonumber\\
0=D_\tau W &\equiv& \partial_\tau W + (\partial_\tau \calK) W = {1\over {\bar
\tau} -\tau} \int_{{\cal M}}
\OL{G}_{(3)} \wedge \Omega\ ,  \label{super}
\end{eqnarray}
where $\chi_\alpha$ is a basis of $(2,1)$ forms on ${\cal M}$.  These
equations imply that $G_{(3)}$ is imaginary self-dual, in correspondence to
the 10d condition (\ref{imsd}).  For F-theory, define a basis
of $(3,1)$ forms $\chi_A$ on $X$; the expected generalization of
(\ref{super}) is
\begin{equation}
0=D_A W = \int_X G_{(4)}\wedge \chi_A\ .
\end{equation}

While our discussion so far has focused on the case where there is only one
K\"ahler modulus, $\rho$, a general model may have several K\"ahler moduli
$\rho_i$.  The required modification of this discussion is quite simple.
The superpotential is independent of all of the $\rho_i$.
It should then follow that the K\"ahler metric for
the K\"ahler deformations produces an analog of the simplification
(\ref{pot}), where now the greek indices sum over moduli excluding the
$\rho_i$.  One way to see this is from the 10d picture -- the condition
(\ref{imsd}), whose correspondence with the 4d potentials was just seen, is
independent of the K\"ahler moduli.
So the no-scale structure survives, with each of the K\"ahler
moduli persisting as a flat direction at this order.
Because it is not difficult to find models with only a
single K\"ahler modulus $\rho$, we will assume that this is the
case in the rest
of the paper.

In the appendix we discuss further the derivation of the
four-dimensional action by dimensional reduction and the correspondence
between the four-dimensional and ten-dimensional pictures.

\sect{Warped solutions and hierarchies}

In section~2 we discussed various global features of IIB compactifications
with a nontrivial warp factor.  We now turn to the local structure of the
warped region.

We
begin by reviewing the solutions of Verlinde \cite{Verlinde}, corresponding
to D3 branes on a compact manifold.  If $N$ D3-branes are coincident, the warp
factor in their vicinity is
\begin{equation}
e^{-4A} \approx \frac{4\pi g_{\rm s} N}{\tilde r^4}\ ,
\label{d3warp}
\end{equation}
with $\tilde r$ the distance from the D3-branes in the $\tilde g_{mn}$
metric. Near the D3-branes the geometry is thus $AdS_5 \times S^5$, producing
a large warp factor~\cite{Verlinde}.
At larger values of $\tilde
r$, the
product structure breaks down due to the curvature of $\calm_6$, and
eventually $\tilde r$ ceases to be a good coordinate~\cite{ver2}:
$\calm_6$ is not globally the product of a five-sphere
and a one dimensional space.  This is similar to the RS2 model~\cite{RS2},
though is a bona-fide compactification, with the compact manifold playing a
role roughly
analogous to the so-called
``Planck brane''  of \cite{RS2}, and yielding a finite
four-dimensional Planck scale.  The warp factor of course diverges as
$\tilde r \to 0$, which is at infinite spatial distance.

If such a model is realized on an orientifold, the dilaton is a constant,
$ e^{\phi} = g_{\rm s}$,
but in the more general context of an
F-theory compactification it varies holomorphically as determined by
(\ref{taueqn}) or equivalently by the eight-dimensional construction.
As we will discuss in section 4.2,
the physics near the D3-branes is essentially the
same, and the
effective value of $g_{\rm s}$ is determined by the value of $\tau$ at
the D3-branes.

To get a large but finite hierarchy, one or more D3-branes must be
separated from the rest by a small distance $\tilde r$.  These might be the
branes on which the Standard Model fields live, or they might be associated
with some symmetry breaking that couples to the Standard Model through the
bulk.  However, the D3-brane coordinates have no potential.
Thus in the present model there is nothing that fixes $\tilde r$ and so the
size of the hierarchy.

In order to find a warped solution that produces a large but
stable hierarchy, we now add fluxes.  Our motivation stems from the
work of Klebanov-Strassler~\cite{KS}.
The basic idea is that locally in the vicinity of a conifold
point, KS have found solutions with fluxes that
generate smooth supergravity solutions with large relative warpings.
Here we will extend this work to the compact context.

CY manifolds are generically nonsingular, but at special values of
the parameters they can develop singularities.  The most generic singular
space is a {\it conifold}~\cite{conifold}.  Locally this can be described
as the submanifold of ${\bf C}^4$ defined by
\begin{equation}
w_1^2 + w_2^2 + w_3^2 + w_4^2 = 0\ . \label{coni}
\end{equation}
This submanifold is singular at
$(w_1,w_2,w_3,w_4) = 0$.  The geometry of this space, including its Calabi-Yau
metric, is described in ref.~\cite{conifold}.  It is important that this is a
good singularity, meaning that string theory makes sense in such a
space~\cite{coni2}.  Although the compactification
space ${\widetilde \calm}_6$ we are using is
either the base of a nontrivial elliptic fibration, or is an orientifold
of a Calabi-Yau, the local structure of a singularity like (\ref{coni})
will not be affected by these global details, so we can use local facts
about CY singularities in the ensuing discussion.

The conifold singularity can be regarded as a cone whose base has the
topology $S^3
\times S^2$.  At the singular point, both the $S^3$ and the $S^2$ shrink to zero
size.  The conifold can be smoothed into a nonsingular CY manifold in two ways.
In the
small resolution of the conifold, the $S^2$ is blown up to finite size.  In
the deformed
conifold,
the $S^3$ is expanded to finite size; it is this that will be relevant for
us.  The
deformed conifold has a simple description as the submanifold
\begin{equation}
w_1^2 + w_2^2 + w_3^2 + w_4^2 = z\ . \label{defcon}
\end{equation}
Here the complex parameter $z$ is the modulus which controls the size
of the $S^3$.

We now consider adding fluxes to this geometry, and find the resulting
potential for the moduli.  Consider a compact manifold with moduli $z$,
$\rho$, and $\tau$ (we explain at the end of this
subsection how additional complex structure moduli
$u_i$ can be incorporated, without substantially modifying the
results).\footnote{More generally, in the case of an F-theory
compactification, the following should be generalized using sections as
outlined in (\ref{Gbundle}), (\ref{Gfdef}).}  Dirac quantization implies that
these fluxes, integrated over all of the three-cycles of the CY, be integers
as in (\ref{fluxquant}). In the vicinity of the conifold, there are two
relevant cycles. Examining the equation (\ref{defcon}),
and taking $z$ to be real and positive for
convenience, the three-cycle which vanishes as $z \to 0$ (denoted $A$) can
be taken to be the $S^3$
on which all of the $w_i$ are real.
In general compact examples, there also exists a dual $B$-cycle which
intersects $A$ exactly once.
An example of such a cycle in this noncompact case
can be constructed by taking $w_{1,2,3}$ to be imaginary and $w_4$ real
and positive.
The KS
solution corresponds to $M$ units of $F_{(3)}$ on the $A$-cycle.  The field
equation in KS
requires that $H_{(3)}$ be supported on the dual cycle
to $F_{(3)}$, so let there be $-K$
units on the $B$-cycle:
\begin{eqnarray}
\frac{1}{2\pi \alpha'} \int_{A} F_{(3)} &=& 2\pi M\ ,\nonumber\\
\frac{1}{2\pi \alpha'} \int_{B} H_{(3)} &=& -2\pi K\ .
\end{eqnarray}
This can also be understood by requiring
D3 charge conservation as in (\ref{threeconst}):
\begin{equation}
{1\over 2\kappa_{10}^2 T_3} \int_{\cal M} H_{(3)} \wedge F_{(3)} = MK \ .
\end{equation}
Thus, in the sense of Poincar\'e duality, we can write
\begin{equation}
F_{(3)} = (2\pi)^2 \alpha' M [B]\ ,\quad
H_{(3)} = (2\pi)^2 \alpha' K [A]\ .
\end{equation}
This gives
\begin{equation}
W = \int_{\cal M} G_{(3)} \wedge \Omega
= (2\pi)^2 \alpha' \biggl( M \int_B \Omega - K \tau \int_A \Omega \biggr)\ .
\label{iibsup}
\end{equation}

The integrals appearing here are the {\it periods} defining the complex
structure of
the conifold.  In particular,
the complex coordinate for the collapsing cycle $A$ is defined by
\begin{equation}
z = \int_{A} \Omega\ .
\end{equation}
It is a standard result that on the dual cycle
\begin{equation}
\int_{B} \Omega \equiv {\cal G}(z)
= \frac{z}{2\pi i} \ln z + {\rm holomorphic}\ .
\label{Bperiod}
\end{equation}
The superpotential is then
\begin{equation}
W= (2\pi)^2 \alpha' (M{\cal G}(z) - K\tau z)\ .
\label{MKpot}
\end{equation}
Such a superpotential has been obtained previously by Vafa~\cite{vatop}.

Let us consider first the $D_z W$ condition,
\begin{equation}
0 = D_zW \propto M\partial_z {\cal G} - K\tau +\partial_z\calK
(M\calG - K\tau z)\ .
\nonumber\\
\end{equation}
In order to obtain a large hierarchy we will take $K/g_{\rm s}$ to be
large: this will result in $z$ being exponentially small.  This has a simple
interpretation in the dual gauge theory, as we will discuss later in this
section.  In this regime, the dominant terms in $D_zW$ are
\begin{equation}
D_z W  \propto \frac{M}{2\pi i} \ln z - i\frac{K}{g_{\rm s}} + O(1)\ ,
\label{dzw}
\end{equation}
It follows that for $K/Mg_s\gg1$, $z$ is indeed exponentially small,
\begin{equation}
z \sim \exp(-2\pi K/Mg_{\rm s})\ .
\label{exphier}
\end{equation}
Thus, we obtain a large hierarchy of scales if, for example, $M=1$ and
$K/g_{\rm s}$ is of order 5.

As things stand, the $D_{\tau}$ equation
\begin{equation}
0=D_\tau W \propto {1\over {\bar\tau}-{ \tau}}(-Kz{\bar \tau} + M \calG)
\label{dtauw}
\end{equation}
cannot be satisfied.
The first term in parentheses is exponentially small, while the second is
not because the holomorphic part in (\ref{Bperiod}) is generically
nonvanishing, $\calG(0) = O(1)$.
Note that this is a property of the compact case.  In the noncompact case
of interest
in KS, the bulk modulus $\tau$ is frozen and there is no corresponding
$D_\tau W$ equation to impose.

The problem arises because at $z=0$ the superpotential (\ref{MKpot}) is
independent of $\tau$, and the remedy is to consider a configuration with
additional $\tau$ dependence.  With such $\tau$ dependence, one can
generically find a solution to (\ref{dtauw}) with $z\approx0$,  though
additional structure may be required to ensure that this minimum is at weak
coupling.   To give one example, $\tau$ can be stabilized by
turning on additional fluxes.
Keeping for simplicity the case of a single complex structure modulus $z$,
there are $2 + 2b_{2,1} = 4$ 3-cycles, namely the pair $(A,B)$ and an
additional pair $(A',B')$.  Turning on $-K'$ units of $H_{(3)}$ on the $B'$
cycle gives
\begin{equation}
W= (2\pi)^2\alpha'\left[M{\cal G}(z) - \tau(K z + K' z') \right]
\end{equation}
where $z'$ is a function of $z$ which is generically nonvanishing at $z=0$,
$z'(z) = O(1)$.  Then if we fix $z = 0$, the $D_\tau W$ equation is
\begin{equation}
0 = D_\tau W \propto
{1\over{\bar \tau} - \tau}[-K'z'(0)\bar\tau + M \calG(0)]\ ,
\label{taustab}
\end{equation}
thus fixing the dilaton at
\begin{equation}
\bar\tau = \frac{M \calG(0)}{K'z'(0)} \ .
\end{equation}
The hierarchy becomes
\begin{equation}
z \sim \exp\biggl( \frac{2\pi K}{K'} \im[{\cal G}(0)/z'(0)] \biggr)
\label{zrat}
\end{equation}
Thus, by appropriate choices of $K$, $K'$, and $M$ one obtains an
exponential hierarchy with the dilaton fixed at an interesting value.

The hierarchy is determined in terms of integer fluxes and the Calabi-Yau
geometry.  To obtain the actual warp factor requires solving the
differential equation~(\ref{warpeq}), but one can estimate it as follows.
The D3-brane warp factor~(\ref{d3warp}) is $e^{4A} \sim \tilde r^4$.  The
resolution of the conifold cuts this off at $w_i^2 \sim z$.  According to
ref.~\cite{conifold}, the conic coordinate $\tilde r$ (which is $\rho$ in the
notation of that paper) is $\tilde r \propto w^{2/3} \propto z^{1/3}$, and so
the hierarchy of energy scales is
\begin{equation}
e^{A_{\rm min}} \sim z^{1/3} \sim \exp(-2\pi K/3Mg_{\rm s})\ .
\label{truehier}
\end{equation}
In effect the fluxes
produce a model similar to
RS1~\cite{RS1}, in which the warp factor does not go
to zero but to a small positive value.\footnote{We should note that, unlike
RS1, there is no negative tension brane at the low energy end; rather,
there is a KS space.  The negative tension objects that we require are
elsewhere on the compact space, in the region that replaces the RS Planck
brane.}

The large hierarchy~(\ref{exphier}) has a simple interpretation in terms of
a dual gauge theory.  The KS solution is the supergravity
dual to a nonconformal ${\cal N}=1$ gauge theory, with confinement and
chiral symmetry breaking at a dynamically generated scale~\cite{KS}.
In the spirit of the Verlinde model~\cite{Verlinde}, the low energy physics
of our supergravity solutions is equivalent to this gauge theory coupled to
the massless bulk fields of the compactification.
The KS solution begins with $N$ whole D3-branes and $M$ fractional D3-branes
at a conifold singularity.  In the end all of these branes are replaced by
flux; their moduli disappear, which is in accord with confinement in the
dual gauge theory.
In particular, with $M$ units of $F_{(3)}$ on the $A$ cycle and $K$
units of $H_{(3)}$
on the dual $B$ cycle, the total D3 charge is $N=MK$.\footnote{In order
to obtain an
interesting low energy spectrum, one may need additional `mobile'
D3-branes in the warped region, but this is beyond our present focus.}

The formula
(\ref{truehier}) then corresponds precisely to the renormalization group
analysis of KS~\cite{KS}.  Using the $\beta$-function in their eq.~(23),
one cascade takes place on a ratio of scales $e^{2\pi /3Mg_{\rm s}}$
(during which the LHS of that equation changes from $-2\pi / g_{\rm s}$ to
$+2\pi/ g_{\rm s}$).  The total number of
cascades is
$N/M = K$, because $M$ units of D3 charge disappear at
each cascade, giving the total hierarchy~(\ref{truehier}).  Thus the
four-dimensional effective action correctly reproduces the physics of the KS
gauge theory.

In the gauge theory, the parameter $z$ is the scale of gluino condensation.
The instability noted in eq.~(\ref{dtauw}) is the familiar fact that a
gluino condensate generates a dilaton potential~\cite{gluino}.  The
stabilization~(\ref{taustab}) does not have a gauge theory origin; rather,
it is a bulk effect in the IIB theory.

There is an effect which might have been expected to destabilize the large
hierarchy, but does not do so.  The dual gauge theory has various relevant
perturbations; for example, the ${\cal N}=1$ supersymmetry allows a
superpotential.  This would produce a mass gap which is of order the
perturbation, rather than exponentially small.  This perturbation is
absent in our solution: in supergravity language it is a 3-form flux, but
it is not of the form ${*}_6 G_{(3)} = iG_{(3)}$, as one sees from the
explicit expressions in section~III.C of ref.~\cite{PS}.  The reason for its
absence appears to be holomorphy: the gauge theory perturbation corresponds
to a growing (nonnormalizable) mode as one move away from the origin, and
evidently this cannot be extended to the full compact space.

So far, we have assumed that there is a single complex structure parameter
$z$.  Suppose there are other complex structure deformations,
controlled by moduli $u_i$.  In such a case, the $u_i$ enter in the
regular terms in the period
(\ref{Bperiod}), so ${\cal G}(z)$ is really ${\cal G}(z,u_i)$.
Generically, assuming that $z$ has been successfully stabilized
near the conifold point in moduli space as above, the equations
\begin{equation}
D_{u_i} W \vert_{z=0} = 0
\end{equation}
can be solved to yield fixed (order 1) values for the other
moduli $u_i$.
So we see that the presence of background RR and NS fluxes generically
serves to fix all of the complex moduli and the dilaton, while leaving
the K\"ahler modulus $\rho$ unfixed.

\sect{Examples}

In order to make our discussion of warped compactifications with fluxes
more explicit and concrete, and in particular check our ability to build
consistent solutions with both negative D3 charge/tension and the above
flux configurations, we now turn to the construction of some explicit
models.  We briefly describe models based on O3-folds, and then
discuss F-theory compactifications in detail.

\subsection{O3 models}

Models in which the negative tension objects are O3 planes are easily
described.  Begin with a CY
manifold with a conifold singularity and a ${\bf Z}_2$ symmetry that has
isolated fixed points, and orientifold on this symmetry.  Since we assuming
that the O3 planes are distant from the singularity, the initial CY must
actually have two conifold singularities which are images of one another.
The D3 charge of the O3 planes is then $-\frac{1}{4}$ times the number of
fixed points.  In order that the supergravity
description be good, we need
$g_{\rm s} N$ to be somewhat greater than one.
To work in perturbative string theory we should also assume
that $g_{\rm s}
\leq 1$.  Therefore, we need $N$, and hence the number of fixed points,
to be large.

We will not present explicit examples, deferring an explicit example to the
discussion of F-theory, but we will present some details of the orientifold
construction and the low energy spectrum.

Let us first determine which of the RR fields survive the orientifolding by
$R\Omega$, where $R$ is the ${\bf Z}_2$ with isolated fixed points, and
$\Omega$ is world-sheet parity.  First, consider a $T^k/{\bf Z}_2$
orientifold, where we can use $T$-duality to relate this to $\Omega$ in the
IIB string~\cite{joesmagnumopus},
\begin{equation}
R\Omega = T^{-1} \Omega T\ .
\end{equation}
Consider a Ramond field with $r$ indices in the direction of the $k$-torus
and $s$
in the orthogonal directions.  In the IIB string, the operator $\Omega$ acts as
$i^{r+s-2}$ on RR potentials and $i^{r+s-3}$ on RR fluxes; thus, for
example, the
RR two-form potential survives the projection to the type I string.  The
$T$-duality takes
$r$ to
$k-r$.  Thus, $\Omega R$ acts as $i^{-r+s+k-2}$ or $i^{-r+s+k-3}$
respectively. We can also phrase this as the statement that the intrinsic
$\Omega R$ of these fields is respectively $i^{n+k-2}$ or $i^{n+k-3}$,
where $n$ is the total number of indices.  This intrinsic parity must be
combined with $(-1)^r$, from the explicit action of the $R$ on the
indices.  For the value
$k=6$ relevant here, the intrinsic parities are respectively $i^n$ and
$i^{n-1}$.

Thus, the Ramond scalar $C$ has even intrinsic parity, as expected because
it is the superpartner of the dilaton.  Similarly $a_{\mu\nu}$, the axionic
part of $\rho$, has even intrinsic parity:
\begin{equation}
C_{\mu\nu pq} = a_{\mu\nu} {\tilde J}_{pq}\ ,
\end{equation}
where ${\tilde J}$ is the K\"ahler form.

The orientifolding requires that the ${\bf Z}_2$ symmetry hold throughout
the moduli space and so only complex structure moduli that are even
survive.  The R-R flux $F_{mnp}$ has odd intrinsic parity, as does the NS-NS
flux $H_{mnp}$ (from the action of $\Omega$).  Thus these must be
proportional to 3-forms of odd intrinsic parity to survive the projection.
Note that the 3-form $\Omega$ (not to be confused with the world-sheet
parity operator) also has odd intrinsic parity.  This is because it is
nowhere vanishing and so in particular is nonzero at the fixed points; at the
fixed points the ${\bf Z}_2$ gives an explicit $-1$ from the indices and
this must be offset by the intrinsic parity.  It follows that the
superpotential
\begin{equation}
\int \Omega \wedge G_{(3)}
\end{equation}
is well-defined on the covering space.  Also, the even complex structure
deformations generate, by contraction with $\Omega$, odd $(2,1)$ forms, so
these are the appropriate fluxes to excite.

Models of this class can be analyzed exactly as in section~3.  One can
choose fluxes through the $A$ and $B$ cycles of the conifold (with the
D3 charge being canceled by the O3 planes), and obtain precisely
the effective field theory for $z, \rho$ and $\tau$ described there.

\subsection{F theory models}

Another general class of warped models arises from F-theory compactifications
to four dimensions.
In such models the
possible configurations of branes and fluxes are constrained by the
topology of the elliptic Calabi-Yau fourfold $X \to \calB$, via
the equation
\begin{equation}
{\chi(X)\over 24} ~=~N_{D3} + {1\over{2\kappa_{10}^2 T_3}}\int_{\calB}
H_{3} \wedge F_{3}\ .
\label{tadpoleF}
\end{equation}
The left-hand side of this equation arises from the induced D3 brane
charge on the wrapped D7 branes, and
this charge must be compensated by introducing either wandering D3
branes or appropriate fluxes in the base ${\calB}$ of the elliptic
fibration.  In general one could also introduce nontrivial gauge bundles
in the wrapped D7 branes (which would yield another term on the right-hand
side of (\ref{tadpoleF}), corresponding to the instanton number in each
D7-brane gauge theory), but we will not need to use this freedom.
Because $\chi \gg 1$ is attainable for Calabi-Yau fourfolds, this class
of models should allow a great deal of freedom in choosing appropriate
flux and brane configurations for model building.
Earlier discussions of fourfold compactifications with nontrivial fluxes
can be found in \cite{beck2,GVW,DGS,greene}.

Because of $SL(2,{\bf Z})$ monodromies around the $(p,q)$ 7-branes
wrapping surfaces in ${\calB}$, the fluxes should really be viewed
as transforming as sections of a nontrivial bundle (as detailed in
section 2.1).  However, we will focus our attention on a local region
around a conifold singularity in the base ${\calB}$, and will write
our formulae in terms of a local trivialization of this bundle.
This is particularly simple in orientifold limits of F-theory vacua,
and we will be most explicit there.  Since the most general F-theory
model does have an orientifold locus in its moduli space \cite{SenO},
this does not constitute a serious loss of generality.

\subsubsection{The Fourfold}

To embed the Klebanov-Strassler system into an F-theory compactification,
we need to exhibit an elliptically fibered Calabi-Yau fourfold $X$ which
admits a conifold singularity in its base ${\calB}$.
A simple example can be designed as follows (the generalization to construct
other examples is straightforward).

Consider for ${\calB}$ the hypersurface given by a quartic equation in
$P^4$
\begin{equation}
P = z_5^2 (\sum_{i=1}^{4} z_i^2) - t^2 z_5^4 +
\sum_{i=1}^{4} z_i^4 = 0
\label{quartic}
\end{equation}
where $z_i$ are the homogeneous coordinates on $P^4$, and $t$ is for
convenience taken
to be a real parameter.
One can construct a fourfold $X$ over ${\calB}$ by specifying a
Weierstrass model (see, {\it e.g.}, \cite{SVW})
\begin{equation}
y^2 = x^3 + x f(z_i) + g(z_i)
\label{weier}
\end{equation}
where $y \in 3L$, $x \in 2L$, $f \in H^{0}(4L)$ and $g \in H^{0}(6L)$; here
$L$ is the line bundle given by
$L = -K_{\calB}$ in terms of the canonical bundle of ${\calB}$.
In practice for this model, we can think of $f$ and $g$ as being
polynomials of degree 4 and 6 in the homogeneous coordinates
$z_i$ (restricted to ${\calB}$).

In type IIB language, one should think of the model (\ref{weier}) as
corresponding to a compactification of IIB string theory on the quartic
in $P^4$, with various $(p,q)$ 7-branes appearing at the loci where
the elliptic fibration degenerates, i.e. where the discriminant
\begin{equation}
\Delta = 4 f^3 + 27 g^2
\label{disc}
\end{equation}
vanishes.
The physics associated with such degenerations involves enhanced
gauge symmetry and more exotic phenomena, and is described for many
cases which arise in compactification on CY threefolds
in \cite{morrvaf,asp,BIKMSV}.
However, for our interests we want a degeneration of the base
which is unrelated to the physics of the 7 branes, and we will simply
insure that the loci in ${\calB}$ of interest to us do not intersect
the $\Delta = 0$ discriminant locus.
For later reference, the value of the IIB axion-dilaton $\tau$ is
determined in terms of the Weierstrass data by the equation
\begin{equation}
j(\tau) ~=~ {4 (24f)^3 \over {4 f^3 + 27 g^2}}
\label{jfun}
\end{equation}
where $j(\tau)$ is the modular invariant function of $\tau$,
normalized so that $j(i) = (24)^3$.

Eqn.~(\ref{Qback}) gives the background D3 charge for this configuration.
For the model (\ref{weier}), one can evaluate
$\chi$ by using the formula in
\cite{SVW}, with the result that
\begin{equation}
-Q_3^{D7}={\chi \over 24} ~=~ 12 + 15 \int_{\calB}
c_1({\calB})^3 ~=~72\ .
\label{Fcharge}
\end{equation}

Inspection of (\ref{quartic}) reveals
that ${\calB}$ has a conifold singularity as
$t \to 0$ -- one can solve $P = dP = 0$ at $(0,0,0,0,1)$.
The collapsing three-cycle
can in fact be exhibited explicitly, as the fixed
point locus of the involution $z_i \to \overline z_i$.
On this locus, the $z_i$ must be real.  One can see from
(\ref{quartic}) that without loss of generality on the real locus
$z_5 \neq 0$,
so we can take $z_5 = 1$ and fix the projective symmetry.
Denote the real part of $z_i$ by $x_i$.
The equation becomes
\begin{equation}
\sum_{i=1}^{4} (x_i^4 + x_i^2) = t^2
\end{equation}
and by defining $u_i = \sqrt{x_i^4 + x_i^2}$, and choosing the branch of
the square root where $sgn(u_i) = sgn(x_i)$, we get a 1-1 map onto the
locus
\begin{equation}
\sum_{i=1}^{4} u_i^2 = t^2
\end{equation}
which describes an $S^3$ that collapses as $t \to 0$.
This is the $A$-cycle of the conifold.

\subsubsection{Orientifold Limit}

Following the work of Sen \cite{SenO}
we can present $X$ on a locus in its moduli space where it has a particularly
simple description, as a type IIB orientifold.
Choose $f$ and $g$ so that they satisfy
\begin{equation}
f = C\eta(z_i) - 3h(z_i)^2,~~g = h(z_i) [C \eta(z_i) - 2h(z_i)^2]
\label{fgspec}
\end{equation}
with $h, \eta$ arbitrary functions of degrees $2$ and $4$.
Since $f$ is quartic this allows for a generic choice of $f$, but
is a specialization of the choice of $g$.
Then from (\ref{jfun}) it is clear that as $C \to 0$ with $\eta$ and
$h$ fixed, $j(\tau) \to \infty$ wherever the numerator does not vanish.
This means $\tau \to i\infty$ almost everywhere on the base, i.e.
we are at weak type IIB coupling.

In fact in this limit, the model becomes an orientifold of type IIB
on a Calabi-Yau threefold
$\Mhat$.
$\Mhat$ is a double cover of ${\calB}$,
specified by the equation (\ref{quartic}) together with
\begin{equation}
\xi^2 - h(z_i)~=~0
\label{mdef}
\end{equation}
where $\xi$ is a new coordinate (valued in the line bundle $L$).
We orientifold $\Mhat$ by the action
\begin{equation}
\xi \to - \xi
\label{orient}
\end{equation}
composed with $\Omega (-1)^{F_L}$
which fixes the locus $\xi = 0$, yielding an O7 plane
localized at $h(z_i) = 0$ in (\ref{quartic}).

One must also introduce $D7$ branes to cancel the RR tadpole generated
by the orientifold. Inspecting the discriminant
$\Delta$, which is
\begin{equation}
\Delta = C^2 \eta^2 (4C \eta - 9h^2)
\end{equation}
in the limit (\ref{fgspec}),
one can see that there are a pair of $D7$ branes located at
$\eta(z_i) = 0$ in $\Mhat$.

\subsubsection{Embedding Klebanov-Strassler}

We have now reduced F-theory on $X$ to IIB string theory on the
orientifold of $\Mhat$ by (\ref{orient}).
Recall that as $t \to 0$, there is a conifold singularity in ${\calB}$,
which survives in the orientifold of IIB on $\Mhat$.
We can choose $h$ and $\eta$ to be of the form
\begin{equation}
h(z_i) = \sum_{i=1}^{5} a_i z_i^2,~~\eta(z_i) = \sum_{i=1}^{5} b_i z_i^4
\label{hetaspec}
\end{equation}
with $a_i$ and $b_i$ real and positive.  With such a choice,
the loci $h=0$ and $\eta=0$ where the $O7$ and $D7$s are located
do not intersect the real slice of ${\calB}$.
But the collapsing three-cycle in $\Mhat$ as
$t^2 \to 0$ lies on this real slice.
Therefore, the $D7$ branes and $O7$ plane do not lie near the
conifold singularity, and we can
work in a local neighborhood of the conifold in the orientifold of
$\Mhat$ while ignoring these other branes.

At the conifold point there is a collapsing $A$ cycle in $\Mhat$,
as well as a
dual $B$ cycle in $\Mhat$ which it intersects once.
We expect to be able to put flux through both of these, consistent with the
orientifold projection.
The background charge (\ref{Fcharge}) is still in force in the orientifold
limit (the $D3$ charge comes from the induced charge on the wrapped
branes); and  can be cancelled
by choosing appropriate $H_3$ and $F_3$
fluxes through these cycles.
If we choose to put $M$ units of RR three-form flux through $A$
and $K$ units of NS three-form flux through $B$, with $MK =N \leq 72$,
then (\ref{tadpoleF}) can be satisfied (for $N < 72$, we
should add wandering D3
branes or turn on other fluxes to saturate (\ref{tadpoleF})).
This allows us to reproduce locally, in a neighborhood of the
conifold point in (the orientifold of) $\Mhat$, the solution of
Klebanov and Strassler \cite{KS}.
That is, the local geometry is the same as the gravity dual of
the $SU(N+M) \times SU(N)$ gauge theory
considered there.
Even with the values of $M$ and $K$ which are possible in this model
(much larger values of $\chi$, and thus larger values of $K$, are possible
in other examples), one can generate a large hierarchy from the
RG cascade, as we have demonstrated in section 3.

Stabilizing the dilaton in such an orientifold
requires some other generic addition to the low-energy superpotential.
One way to accomplish this is to turn on additional fluxes, as
discussed in \S3.
An alternative is to work at generic points in the F-theory moduli
space, which we discuss below.

\subsubsection{Deforming Away from the Orientifold Limit}

To understand the low energy physics governing an orientifold model
with a conifold singularity and appropriate fluxes,
one should compute
the effective field theory governing (at least) three different moduli,
as described in \S3.  These are the complex modulus $z$ which controls
the volume of the collapsing three-cycle at the conifold, the
dilaton $\tau$, and the overall volume of the space $\rho$.

In our F-theory situation, however, we could consider moving away
from the limit of \S4.2.2, so that the model is not a perturbative
IIB orientifold.
Working away from the orientifold limit while keeping
the $F_3$ and $H_3$ fluxes as before, one achieves some
simplification.\footnote{Note that fluxes which were projected in
by the orientifold action are guaranteed to adiabatically
deform to consistent $G_{(4)}$ fluxes in the full CY fourfold geometry.}
While $\rho$ (the size of the base ${\calB}$)
and $z$ (here controlled by $t^2$ in (\ref{quartic})) remain moduli
in the F-theory picture, the dilaton does not remain an independent
modulus.  It is fixed in terms of the complex structure of $X$ by
the equation (\ref{taueqn}), with solution
(\ref{jfun}).

This means that the naive problem with solving the $D_{\tau}W = 0$ equation
in the vicinity of the conifold point, solved in \S3\ by for example
turning on
an additional flux, will not occur here.  $\tau$ does not appear as an
independent mode in the low-energy effective field theory.  The modes
controlling the complex structure of $X$, which determine $\tau$
via (\ref{jfun}), are frozen on general grounds by just the Klebanov-Strassler
fluxes, as described at the end of \S3.
Although our discussion there was in terms of perturbative type IIB string
theory, there is an alternative derivation which goes through M-theory.
One can view F-theory on $X$ as being defined by a
limit of M-theory on $X$
(where one shrinks the volume of the elliptic fiber in going from M-theory
to F-theory).  The superpotential for complex structure moduli in
M-theory on $X$ is given by the formula
%
(\ref{Fsuper})
where $G_{(4)}$ is the M-theory four-form flux and $\Omega_{4}$ is
the holomorphic (4,0) form on $X$.
The formula (\ref{iibsup}) for the IIB string theory superpotential
follows from (\ref{Fsuper}) in the F-theory limit, for
suitable choices of $G_{(4)}$ (those which lift to $G_{(3)}$ flux in IIB
language) and in the case that $X$ is a Calabi-Yau threefold times a
two-torus.
In the more general F-theory case, $X$ is not such a product,
but nevertheless the $A$ and $B$ cycle in ${\calB}$ that we have been
using lift to 4-cycles in $X$ and allow use of the local decomposition
(\ref{Gfdef}).  The statement that the complex moduli
(and therefore the value of $\tau$ at the conifold point in ${\calB}$)
are fixed then follows from the fact that the period of $\Omega_{4}$
over the lift of the $B$ cycle will have generic dependence on the
complex structure moduli.

We saw in \S3\ that fixing the
dilaton, either by this mechanism or by turning on additional fluxes,
allows one to solve for $z$.
The exponentially small value of $z$ computed from the superpotential
of \cite{vatop} independently confirms the existence of a hierarchy
for reasonable choices of $M$ and $N$
(and represents the small, dynamically generated scale of chiral
symmetry breaking in \cite{KS}).

\sect{Conclusion}

There has been a great deal of interest in
finding string theory constructions which produce large hierarchies through
warping, and in particular reproduce, at long wavelengths,
features of the RS1 model~\cite{RS1}.
Building on the ideas of Verlinde and collaborators \cite{Verlinde,ver2},
we have described orientifold and F-theory models which accomplish this.
The role of the AdS throat and the infrared brane
is played by (a finite radial segment of)
the gravity dual to a confining gauge theory found by Klebanov and
Strassler \cite{KS}, while the UV brane is replaced by
the bulk of the string theory compactification manifold.

Our models are consistent, nonsingular string theory backgrounds.
However, we expect $\alpha^\prime$ and
string loop corrections to generate a potential for the
overall scale $\rho$ of the compactification manifold.  An analogous
problem also arises in familiar classical heterotic string backgrounds
\cite{witeff,gluino}, and in some ways our models are quite similar to
those (with the important difference that non-perturbative gauge theory
effects have already been incorporated in the classical gravity solution).
It would be very interesting to find mechanisms for stabilizing $\rho$ in
these models; toy models where all of the moduli are stabilized by fluxes
can be constructed \cite{toappear}.

The duality between gauge theories and compactifications with flux
extends beyond the single example \cite{KS}
we have used here.  The results of \cite{vatop} provide a
more general construction of dualities between fluxes and gauge
theories, and quantum gauge theory effects are again
calculable using classical geometry.
It would be interesting to use other examples of this gauge theory/flux
duality to construct ${\cal N}=1$ string compactifications with
moduli which are calculably stabilized by non-perturbative gauge dynamics.

Finally, it has recently become clear that warped compactifications
offer new mechanisms, distinct from AdS redshifting, of producing
large hierarchies \cite{dkkls}.
The relevant warped models need to have two or more different brane
throats, with fairly generic warping (so power-law warping
is sufficient).  Large hierarchies can then be produced by the
tunneling-suppressed (and therefore weak) interactions
between the IR modes localized down distinct throats.
It should be possible to design string theory examples of such
multi-throat compactifications by
generalizing the construction in our paper.

\subsection*{Acknowledgments}

We would like to thank P. Aspinwall, A. Grassi, S. Gubser, S. Gukov, G.
Horowitz, S. Katz, I. Klebanov, 
M. Schulz, E. Silverstein, C. Vafa, H. Verlinde and E. Witten for
helpful discussions.
This work was supported by National Science Foundation
grants PHY99-07949 and PHY97-22022, and by the
DOE under contracts DE-FG-03-91ER40618 and
DE-AC03-76SF00515.
The research of S.K. is supported in part by a
David and Lucile Packard Foundation Fellowship for Science and Engineering
and an Alfred P. Sloan Foundation Fellowship.

\appendix
\sect{Dimensional reduction}

We now develop further the low energy effective action, discussed in
section~2.   Before turning on fluxes, the underlying manifold ${\widetilde
{\cal M}}$
generically has a large number of moduli and corresponding massless
supermultiplets in the four-dimensional low-energy effective theory.
Turning on fluxes deforms the geometry of the compactification, and in the
four-dimensional effective  theory generates a potential for the massless
moduli~\cite{GVW,DGS}.

\subsection{Kinetic terms and K\"ahler potential}

The allowed moduli depend on the topology of the compactification, though
one generically has the universal
K\"ahler modulus corresponding to overall rescaling
of the six-dimensional metric.  This has partner $a_{\mu\nu}$, arising from
\begin{equation}
C_{\mu\nu p q} = a_{\mu\nu} {\tilde J}_{pq}
\label{axion}
\end{equation}
where ${\tilde J}$ is the K\"ahler form.
We work in the approximation of constant warp factor and vanishing $\tilde
F_5$; as discussed in section 2.2.4 this is valid in the large-radius limit
(although we expect our
expressions to generalize beyond this).
The effective action for this K\"ahler multiplet together with
the 4d metric and dilaton can be found by computing the action
(\ref{IIBE}) with
\begin{equation}
ds^2 = g_{\mu\nu}dx^\mu dx^\nu + e^{2u(x)} {\tilde g}_{mn} dy^m dy^n
\end{equation}
where ${\tilde g}_{mn}$ is
the metric of the compactification.
In doing so, we define the 4d Newton's constant $\kappa_{4}^2 =
\kappa_{10}^2 / \tilde V$ where $\tilde V$ is the volume of
$\widetilde {\cal M}$, and the Weyl rescaled metric
$g_4 = e^{-6u } \tilde g_4$.  We also dualize,
$d a_{(2)} = e^{ - 8u}{\tilde \ast} db$, and define $\rho=b/\sqrt{2}+
ie^{4u}$.
The result is
\begin{equation}
S= \frac{1}{2\kappa_{4}^2} \int\! d^{4}x\,
(-\tilde g_4)^{1/2} \Biggl( \tilde R_4 - 2 \frac{\partial_\mu \tau \partial^\mu
\bar\tau}{|\tau - \bar\tau|^2} - 6 \frac{\partial_\mu \rho \partial^\mu
\bar\rho}{|\rho - \bar\rho|^2} \Biggl)
 \ .
\end{equation}
The kinetic terms for $\tau$ and $\rho$ can thus be found from the K\"ahler
potential
\begin{equation}
\calK_1 = -\ln[-i(\tau-\bar\tau)] - 3\ln[-i(\rho-\bar\rho)]\ .
\label{konepot}
\end{equation}

In the O3 case, both $\tau$ and $\rho$ survive the
projection.  In the case of an F-theory compactification, the D7-brane
monodromies generally remove $\tau$ from the 4d spectrum, although $\tau$
varies as other complex structure moduli, e.g. parameterizing the locations
of the D7 branes, vary.

The remaining moduli are the other
K\"ahler and complex structure deformations of the
6d compactification, or, in the F-theory context, of the eight-dimensional
Calabi-Yau manifold.
In the following, we imagine for definiteness that $\widetilde\calm$ is a
Calabi-Yau orientifold, and we discuss the complex structure moduli
space of Calabi-Yau threefolds, but the relevant parts of the story
carry over also to the F-theory examples.

 As shown by Candelas and de la Ossa~\cite{modsp}, the
effective action for CY moduli
is determined by the Weil-Petersson metric on the moduli
space, and one may derive a simple expression for the corresponding K\"ahler
potential.
First note that on a general CY threefold there are the following harmonic
forms:
\begin{enumerate}
\item
One (3,0) form $\Omega$.
\item
$b_{2,1}$ primitive (2,1) forms $\chi_\alpha$.
\item
Their (1,2) conjugates $\OL{\chi_\alpha}$.
\item
The (0,3) conjugate $\OL\Omega$.
\end{enumerate}
These satisfy
\begin{equation}
{*_6} \Omega = -i \Omega\ ,\quad
{*_6} \chi_\alpha =  i \chi_\alpha\ .
\label{Selfdual}
\end{equation}
As discussed in section~4.2, only forms of odd intrinsic parity under the ${\bf
Z}_2$ projection are relevant.  This includes $\Omega$ and a subset of the
$\chi_\alpha$.  In the subsequent analysis $\alpha$ is restricted to this
subset.

The metric for the complex structure deformations takes the form
\begin{equation}
G_{\alpha\bar\beta} = - \frac{\displaystyle\int_{{\cal M}} \chi_\alpha
\wedge
\OL{\chi_\beta} } {\displaystyle\int_{{\cal M}} \Omega \wedge \OL{\Omega}
}\ .
\label{WPmet}
\end{equation}
To find the corresponding K\"ahler potential, let
$z^\alpha$ be coordinates on the complex structure
moduli space.  Then it can be shown that
$\partial \Omega/\partial z^\alpha$ is
$(3,0) + (2,1)$, and more precisely that there is a basis $\chi_\alpha$
such that
\begin{equation}
\frac{\partial \Omega}{\partial z^\alpha} = k_\alpha(z,\bar z) \Omega +
\chi_\alpha\ .
\label{kdefn}
\end{equation}
Defining
\begin{equation}
\calK_2 = - \ln\biggl(-i \int_{{\cal M}} \Omega\wedge \OL{\Omega} \biggr)\
,
\label{Ktwodef}
\end{equation}
one may then show
\begin{equation}
\partial_\alpha \calK_2 = -k_\alpha\ ,
\label{kadef}
\end{equation}
and the equation
\begin{equation}
\partial_\alpha \partial_{\bar\beta} \calK_2
= G_{\alpha\bar\beta}\
\label{KtoM}
\end{equation}
gives the above metric.

In the context of an F-theory compactification, an obvious generalization
of (\ref{Ktwodef}) is
\begin{equation}
\calK =  - \ln\biggl(\int_{{X}} \Omega_4\wedge \bar \Omega_4 \biggr)\ .
\end{equation}

\subsection{The potential and superpotential}

We now turn to the problem of finding the potential determined by the
fluxes.
{}From (\ref{IIBE}), the potential is determined by
\begin{equation}
S_G=-{1\over 24\kappa_{10}^2} \int_{\calm} d^6y {\tilde g}^{1/2}
{G_{mnp} {\bar G}^{\widetilde{mnp}} \over \im\,\tau}\ .
\end{equation}
Again, we are in a large-radius approximation where the warp factor is
constant and $\tilde F_{(5)}=0$.
We define the imaginary self-dual parts of $G_{(3)}$ as
\begin{eqnarray}
G_{(3)} &=& G_{(3)}^+ + G_{(3)}^-\ ,\quad
G_{(3)}^{\pm} = \frac{1}{2} ( G_{(3)} \pm i {*_{6}} G_{(3)}) \ ,\nonumber\\
{*_{6}} G_{(3)}^\pm &=& \mp i G_{(3)}^\pm\ .
\end{eqnarray}
The action can then be written as
\begin{eqnarray}
S_G &=& -\frac{1}{12\kappa_{10}^2 {\rm Im}\, \tau}
\int_{{\cal M}} d^6x \,\tilde g^{1/2}
G^+_{mnp}
\OL{G}^{+\widetilde{mnp}}-\frac{i}{4\kappa_{10}^2 {\rm Im}\,
\tau } \int_{{\cal M}} G_{(3)}
\wedge
\OL{G}_{(3)}
\nonumber\\
&=&
-{\cal V}
-\frac{i}{4\kappa_{10}^2 {\rm Im}\, \tau} \int_{{\cal M}} G_{(3)}
\wedge
\OL{G}_{(3)}
 \ \label{Gact}
\end{eqnarray}
where we define the potential
\begin{equation}
{\cal V} =-\frac{1}{2\kappa_{10}^2 {\rm Im}\, \tau}
\int_{{\cal M}}  G_{(3)}^+ \wedge
{*_{6}} \OL{G}_{(3)}^+
\end{equation}
The second term in (\ref{Gact})
is proportional to
$\mu_3 Q^G_3$, where $\mu_3$ is the D3 tension and $Q_3^G$
is the D3
charge carried by the three-form flux.  This term is topological
and does not involve the moduli.  It is canceled by the tension of the
localized sources, because these have total D3 charge $Q_3^{\rm loc}
= - Q_3^G$ and saturate the inequality~(\ref{bps?}).

(\ref{Selfdual}) implies that ${\cal V}$ only depends on the
coefficients of $\Omega$ and $\OL{\chi_\alpha}$ when $G_{(3)}$
is expanded in the basis of 3-forms.  
In terms of the metric (\ref{WPmet}), we
find
\begin{equation}
{\cal V} =
\frac{\displaystyle
i\int_{{\cal M}} G_{(3)} \wedge \OL{\Omega}
\int_{{\cal M}} \OL{ G}_{(3)} \wedge {\Omega} + G^{\alpha\bar\beta}
\int_{{\cal M}} G_{(3)} \wedge \chi_\alpha
\int_{{\cal M}} \OL{ G}_{(3)} \wedge \OL{\chi_\beta}
}{2 {\rm Im}\, \tau\, \kappa_{10}^2 \displaystyle\int_{{\cal M}} \Omega
\wedge \OL{\Omega} } \ .
\end{equation}
This can be derived from a superpotential of the form discussed in
refs.~\cite{GVW,vaftay,vatop},
\begin{equation}
W= \int_{\calm} G_{(3)} \wedge \Omega\ .
\label{superpot}
\end{equation}
Indeed, from (\ref{kdefn},\ref{konepot})
we find
\begin{eqnarray}
D_\alpha W &\equiv& \partial_\alpha W + (\partial_\alpha 
\calK) W = \int_{\cal
M} G_{(3)}
\wedge
\chi_\alpha \ ,\nonumber\\
D_\tau W &\equiv& \partial_\tau W + (\partial_\tau \calK) W ={1\over {\bar
\tau}-\tau}  \int_{{\cal M}}
\OL{G}_{(3)} \wedge \Omega\ ,  \label{su}
\end{eqnarray}
where $\calK = \calK_1 + \calK_2$.
After a Weyl transformation to the four-dimensional Einstein frame, the
potential takes the standard ${\cal N}=1$ supergravity
form \cite{n1pot}, as in eq.~(\ref{pot}).

This potential has been discussed before~\cite{previous}, but in somewhat
different contexts.  In the first place, these earlier systems had $\CN=2$
low energy supersymmetry, even when the potential was written in $\CN=1$
form.  Here, the orientifolding or the F-theory D7 configuration
explicitly reduces the low energy
supersymmetry to $\CN=1$.
Second, objects with negative D3 charge were not included, so the fluxes
were restricted to $\int_{\cal M} H_{(3)} \wedge F_{(3)} = 0$.

The conditions $D_\alpha W= D_\tau W = 0$ imply that $G^+_{(3)} = 0$.  Thus
the effective four-dimensional action reproduces the ten-dimensional
conditions (\ref{imsd})
for a solution.  Unbroken supersymmetry requires also that
$D_\rho W = 0$, implying that the $(0,3)$ part of $G_{(3)}$ vanishes and so
this flux is $(2,1)$ and primitive, again as argued directly in ten
dimensions.  The latter condition is equivalent to $W=0$; this will
generically not hold when $D_\alpha W= D_\tau W = 0$.

The F-theory generalization of this discussion readily follows, with
superpotential \cite{GVW}
\begin{equation}
W=\int_X G_{(4)}\wedge\Omega_4\ ,
\end{equation}
where $G_{(4)}$ is the F-theory lift of the flux, locally given in
eq.~(\ref{Gfdef}).

This dimensional reduction has been carried out in a limit that is rather
orthogonal to the main concerns of this paper, in that the warp factor is
constant rather than strongly varying, and $\tilde F_{(5)}=0$.  The
detailed treatment of dimensional reduction in the warped case is left for
the future (see also ref.~\cite{greene}), but in the present case we can
argue that the key results are unaffected. In particular, the
ten-dimensional analysis of section~2 shows that the solutions found from
the effective action derived here remain solutions even when the warping is
taken into account.  The physical reason is that all localized sources as
well as the supergravity fields couple to the warp factor and the 5-form
flux in the same ratio, so that there is no net force.

The superpotential derived in the large-radius limit is exact in string
perturbation theory.  This is because the real part of $\rho$ is an axion,
obtained from the tensor field~(\ref{axion}), and so there is a
Peccei-Quinn symmetry broken only by D-instanton effects.  Thus $\rho$
cannot appear in the superpotential~\cite{newiss}; the same will be true of
all other K\"ahler moduli.  Note that this is not true of $\tau$.  The
field $C_{(0)}$ appears in the classical action through the definition of
$G_{(3)}$, so there is no PQ symmetry and $\tau$ does enter into the
classical superpotential~(\ref{superpot}).

\end{document}